\documentclass[prl,showpacs,byrevtex]{revtex4}
\input epsf.tex
\def\DESepsf(#1 width #2){\epsfxsize=#2 \epsfbox{#1}}
\usepackage{graphicx}
\usepackage{epsfig}
\usepackage{bm}
\def\ba{\begin{eqnarray}}
\def\ea{\end{eqnarray}}
\def\br{\begin{array}}
\def\er{\end{array}}
\def\be{\begin{equation}}
\def\ee{\end{equation}}
\def\uoo{c_{13}c_{12}}
\def\uot{c_{13}s_{12}}
\def\uoth{s_{13}e^{-i\delta}}
\def\uto{-c_{23}s_{12}-c_{12}s_{13}s_{23}e^{i\delta}}
\def\utt{c_{12}c_{23}-s_{12}s_{13}s_{23} e^{i\delta}}
\def\utth{c_{13}s_{23}}
\def\utho{s_{12}s_{23}-c_{12}s_{13}c_{23}e^{i\delta}}
\def\utht{-c_{12}s_{23}-c_{23}s_{13}s_{12} e^{i\delta}}
\def\uthth{c_{13}c_{23}}
\def\ao{\alpha_1}
\def\at{\alpha_2}
\def\d{\delta}

\def\Z{M_Z}

\newcommand\bi{\begin{itemize}}
\newcommand\ei{\end{itemize}}

\def\Dtho{\Delta {m_{31}^2}}
\def\Dto{\Delta {m_{21}^2}}
\def\Dtht{\Delta {m_{32}^2}}

\def\sc{\sin\theta_{13}}
\def\sa{\sin\theta_{23}}

\def\dsol{ds_{12}\over dt}
\def\dat {ds_{23}\over dt}
\def\dch{ds_{13}\over dt}

\def\da1{d\alpha_1\over dt}
\def\da2{d\alpha_2\over dt}
\begin{document}
\thispagestyle{empty}
\title{ Neutrino Mixings  and  Leptonic CP Violation from CKM
 Matrix and Majorana Phases}
\author{Sanjib Kumar Agarwalla}
\email{sanjib@mri.ernet.in}
\affiliation{Harish-Chandra Research Institute, Chhatnag Road, Jhunsi, Allahabad 211019, India}
\author{M. K. Parida}
\email{paridam@mri.ernet.in}
\affiliation{Harish-Chandra Research Institute, Chhatnag Road, Jhunsi, Allahabad 211019, India}
\author{R. N. Mohapatra}
\email{rmohapat@physics.umd.edu} \affiliation{Department of
Physics, University of Maryland, College Park, MD 20742, USA.}
\author{G. Rajasekaran}
\email{graj@imsc.res.in}
\affiliation{Institute of Mathematical Sciences, Chennai 600113, India.}
\begin{abstract}
The high scale mixing unification hypothesis recently proposed by
three of us (R. N. M., M. K. P. and G. R.) states that if at the
seesaw scale, the quark and lepton mixing matrices are equal then
for quasi-degenerate neutrinos, radiative corrections can lead to
large solar and atmospheric mixings and small reactor angle at the
weak scale in agreement with data. Evidence for quasi-degenerate
neutrinos could, within this framework, be interpreted as being
consistent with
quark-lepton unification at high scale. In the current work, we
extend this model to show that the hypothesis works quite
successfully in the presence of CP violating phases (which were
set to zero in the first paper). In the case where the PMNS matrix
is identical to the CKM matrix at the seesaw scale, with a Dirac
phase but no Majorana phase, the low energy Dirac phase is
predicted to be ($\simeq 0.3^{\circ}$) and leptonic CP-violation
parameter $J_{CP} \simeq (4 - 8 )\times 10^{-5}$  and $\theta_{13}
= 3.5^{\circ}$. If on the other hand, the PMNS matrix is assumed
to also have non-negligible Majorana phase(s) initially, the resulting 
theory damps
radiative magnification phenomenon for a large range of parameters
but nevertheless has enough parameter space to give the two
necessary large neutrino mixing angles. In this case, one has
$\theta_{13} = 3.5^{\circ} - 10^{\circ}$ and $|J_{CP}|$ as large
as $0.02-0.04$ which are accessible to long baseline neutrino
oscillation experiments.
\end{abstract}

\date{\today}

\rightline{hep-ph/0611225}
\rightline{HRI-P-06-11-004}
 \maketitle

\section{I~Introduction}

Grand unified theories \cite{ps, georgi, so10, rev} with quark-lepton
unification have often been used as  key  ingredients in attempts
to understand the widely differing values of physical parameters
describing particle interactions at low energies. In the context
of SUSY GUTs, this approach can explain the experimentally
measured values of the electro-weak mixing angle. The same models
also lead to $b-\tau$ Yukawa unification \cite{buras} which seems
to be in rough agreement with observation or even $t-b-\tau$
Yukawa unification \cite{ananth} which agree with observation for
large values of $\tan \beta (= v_u/v_d)$. It is then natural to
explore whether there are other manifestations of quark lepton
unification at low energies.
\par
In a recent paper, three of us \cite{mpr1} discussed the
possibility that weak interaction properties of quarks and leptons
parameterized by very different flavor mixing matrices at low
energies may become identical at high scales and provide another
signature of quark--lepton unification. We found that if neutrinos
are Majorana fermions with quasi-degenerate masses and with same
CP, it could indeed happen i.e. starting with the CKM mixing
matrix for neutrinos at the GUT-seesaw scale \cite{seesaw}, as
would be expected on the basis of quark--lepton unification \cite{ps},
renormalization group evolution (RGE) to the weak scale
leads to predictions for neutrino mixings in agreement with
observations \cite{maltoni}. Since small angles become larger, we
have called this phenomenon radiative magnification and the
interesting result is that the RGEs give two large mixings in the
solar and atmospheric neutrino sectors while ultra smallness of
$V_{ub}$ even after radiative magnification yields a small
$\theta_{13}$ maintaining consistency with CHOOZ-Palo-Verde
observations \cite{chooz}. We then predicted $\sin\theta_{13} \le
0.08-0.10$ which could be used to test the model \cite{mpr1}. We called this
``high scale mixing unification'' (HUM) hypothesis. The common
mass of quasi-degenerate neutrinos required for HUM to work is in
the range $0.15$ eV $\le {\rm m}_i \le 0.65$ eV. This falls in the
appropriate range accessible to the currently planned
experiments \cite{bb} and overlaps with the values claimed by
Heidelberg-Moscow ${\beta\beta}_{0\nu}$ experiment \cite{klapdor}.
It also overlaps the WMAP bound \cite{wmap} and the range
accessible to the KATRIN experiment \cite{katrin}.
It is also interesting to note that the 
radiative magnification mechanism
with high-scale mixing 
unification works
only for reasonably large values of $\tan\beta$ for 
which $t-b-\tau$ Yukawa
unification takes place.

Furthermore, the HUM hypothesis provides an alternative way to
understand the difficult problem of the diverse mixing patterns
between quarks and leptons without relying on new mass textures
for neutrinos or new family symmetries. This makes it interesting
in its own right and in our opinion deserves further
consideration.

\par
Renormalization group effects on neutrino mass parameters over
wide range of values have been discussed in a number of works
using standard \cite{wett, babu, casas, bdmp, antusch1, antusch2, ellis} or
nonstandard \cite{xing} parameterizations of the PMNS matrix; but
our present and earlier results \cite{mpr1, mpr2, mpr3} are distinct
in the following respects: (1) because of the underlying
quark-lepton unification hypothesis, the input values of the three
mixing angles are small  at the GUT-seesaw scale and are identical
to the corresponding CKM mixings \cite{mpr1}. The high scale input
value of the Dirac phase in the PMNS matrix, which was set to zero
in \cite{mpr1} and discussed in subsequent sections is also
identical to the corresponding quantity in the CKM matrix. (2) The
three light neutrinos are quasi-degenerate in mass. (3) The RH
Majorana neutrino mass
 matrix is proportional to a unit matrix
due to an $S_4$ \cite{mpr1, lee} symmetry, which leads to
quasi-degenerate light neutrinos  through the type-II seesaw
mechanism. (4) The heavy degenerate masses of RH neutrinos being
very close to the the GUT-seesaw scale, large threshold
corrections originating from non-degeneracy of RH neutrinos are
absent in our case. In view of this, the model predictions are
definite and can be tested more by planned neutrino experiments.

\par
         For the sake of simplicity, only CP conserving case
was considered in \cite{mpr1} and all neutrino mass eigenstates were
assumed to possess the same CP. Clearly, understanding CP phases
is an integral part of understanding the flavor puzzle \cite{smir}
and it is therefore important to explore whether the ``high scale
mixing unification'' hypothesis throws any light on this. The
purpose of the present paper is to investigate this question.

      Following the HUM hypothesis we
identify the three mixing angles and the Dirac phase of the PMNS
mixing matrix as those of the CKM mixing matrix at the GUT-seesaw
scale while keeping the two Majorana phases as unknown. Whether
exact quark-lepton symmetry at high scale permits nonzero values
of the Majorana phases is a model dependent question; therefore,
we consider two different cases: (i)  when the high scale PMNS
matrix with zero Majorana phases is identified with the CKM
matrix; and (ii)  when the PMNS matrix equals the product of the
CKM matrix times a diagonal Majorana phase matrix. The
diagonal-phase matrix  may originate from the seesaw contribution
to the neutrino mass.

We find that the  radiative magnification of mixing angles are
generally damped in the presence of Majorana phases and the degree
of damping depends upon the  values of these parameters. However
in spite of this, the model has enough parameter space to magnify
the mixing angles to be in agreement with   the low-energy
neutrino data starting from the CKM matrix at the GUT-seesaw
scale; in this case, we find that $\theta_{13}$ and leptonic
CP-violations are large enough to be measurable in long baseline
neutrino oscillation experiments.

We also  derive new analytic formulas for threshold corrections
including Majorana phases which are generally valid in all models
with quasi-degenerate neutrino masses.
 Out of a number of solutions which require ~small
threshold corrections to bring the solar neutrino  mass squared
difference in accord  with experimental data as before, we find a
new region in the parameter space where no such threshold
corrections are necessary. We show how the the partial damping of
magnification due to Majorana phases  is utilized  to obtain this
new class of solutions.

\par

This paper is organized as follows. In Sec.II we discuss the new
boundary condition at the GUT seesaw scale and RGEs. In Sec.III we
discuss input parameters at GUT-seesaw scale and some general
criteria including the role played by Majorana phases  in damping
radiative magnification.
In Sec.IV we present new analytic formulas showing how Majorana
phases influence threshold corrections explicitly.
Solutions to the RGEs and predictions in
different cases are presented in Sec.V. In Sec.VI we discuss the
new class of solutions without threshold corrections using damping
property of Majorana phases. In Sec.VII we summarize the results
briefly and state our conclusions.

\section{II~ RG Equations and High Scale Boundary Condition }

The starting point of our analysis is the assumption that there is
an underlying gauge symmetry  that unifies quarks and leptons and
lead to the possibility that the CKM matrix for quarks and the
PMNS matrix are equal. In \cite{mpr1}, we gave an example of this
symmetry as $SU(2)_L\times SU(2)_R \times SU(4)_C$ times a discrete
permutation family symmetry $S_4$. In this paper, we do not discuss any
specific model but rather assume that such a symmetry exists
implying the mixing unification at the seesaw scale. Given this
hypothesis, the existence of the CKM phase would imply that the
Dirac phase of the PMNS matrix at the GUT scale is the CKM phase.
As far as the two Majorana phases of the PMNS matrix are
concerned, they have no counterparts in the CKM matrix. However,
due to seesaw mechanism, it is not implausible that the Majorana
phases are also present at the high scale. We therefore study two
possibilities: (i) setting the two Majorana phases to zero at the
GUT-seesaw scale and examine the low energy behavior of the
theory; (ii) treating the two phases as unknown parameters and see
their effect.

The boundary condition for mixing angles at
the seesaw scale is therefore given by :\\
\noindent {\bf $\mu = M_R$:} \be U_{PMNS}(M_R) = V_{CKM}(M_R)
\times V_D(M_R)  \label{eq1} \ee \noindent where $V_{CKM}$ is the
RG-extrapolated CKM matrix for quark mixings from low energy data
including its experimental value for the Dirac phase.  The two
unknown Majorana phases are provided through the diagonal matrix
$V_D={\rm {diag}}( e^{i\alpha_1 }, e^{i\alpha_2 },1)$. Thus, up to
the presence of the diagonal phase matrix in eq.(1), this boundary
condition establishes complete identity of the PMNS mixing matrix
for leptons with the CKM mixing for quarks  at the GUT-seesaw
scale through the high scale mixing unification (HUM) hypothesis.
Below the GUT-seesaw scale, however, there could be substantial
differences between the two due to RG evolution effects
and we parameterize  the PMNS matrix as\\
\noindent {\bf $\mu < M_R$ :}\\
\be U=\left[\br{ccc}
\uoo&\uot&\uoth\\
\uto&\utt&\utth\\
\utho&\utht&\uthth
\er\right]\times {\rm {diag}}.\left(e^{i\alpha_1 }, e^{i\alpha_2 },1\right)
 \label{eq2}\ee
\par
\noindent where $c_{ij}=\cos\theta_{ij}$ and
$s_{ij}=\sin\theta_{ij} (i,j=1, 2, 3$) and all the mixing angles
and phases are now scale dependent. We have chosen the Majorana
phases to be two times of that in usual parameterizations. This
choice is made for the sake of convenience in computation.
The values of the mixing angles at low
energies will be obtained from RG evolution
following the top-down approach under the boundary conditions,
\par
\noindent
\ba
\sin\theta_{12}^0 & \equiv & \sin\theta_{12} (M_R)_{PMNS} = \sin\theta_{12} (M_R)_{CKM} , \nonumber\\
\sin\theta_{23}^0 & \equiv & \sin\theta_{23} (M_R)_{PMNS} = \sin\theta_{23} (M_R)_{CKM} , \nonumber\\
\sin\theta_{13}^0 & \equiv & \sin\theta_{13} (M_R)_{PMNS} = \sin\theta_{13} (M_R)_{CKM} , \nonumber\\
\delta^0 & \equiv & \delta (M_R)_{PMNS} = \delta (M_R)_{CKM} \label{eq3}
\ea

\par
\noindent and the two Majorana phases at $\mu = M_R$ will be
treated as unknown parameters, $\ao^0$ and $\at^0$. 
The renormalization group equations
(RGEs) for the neutrino mass matrix in the flavor basis were derived
earlier and have been used to obtain  a number of interesting
conclusions especially for quasi-degenerate
neutrinos \cite{babu, bdmp} in the flavor basis. As in the earlier
case we find here more convenient to use RGEs directly for the
mass eigenvalues and mixing angles including phases derived in the
mass basis \cite{casas, antusch2}. For the sake of RG solutions  assume
the three neutrino
mass eigenvalues to be real and positive.
 The RGEs for the light neutrino mass eigenvalues
 can be written
as
\par
\noindent
\be
{dm_i\over dt}=-2F_{\tau}(P_i + G_i)m_i -
m_iF_u,\,\left(i=1,2,3\right) \label{eq4}
\ee
\par

The RGEs for  mixing angles and phases are 
represented upto a good approximation
for small $\theta_{13}$ which is actually the 
case through the following equations \cite{antusch2},

\ba
\dat&=&\frac{F_{\tau}c_{23}\sin2\theta_{23}}{2(m_3^2-m_2^2)}\left[c_{12}^2
 (m_3^2+m_2^2+2m_3m_2\cos 2\alpha_2)\right.\nonumber \\
&&\left.+ s_{12}^2(m_3^2+m_1^2+2m_3m_1\cos 2\alpha_1)/(1+R)\right],
\label{eq5}\\
\dch&=&-\frac{F_{\tau}c_{13}\sin2\theta_{12}\sin 2\theta_{23}m_3}
{2(m_3^2-m_1^2)}\left[m_1\cos(2\alpha_1-\delta)\right.\nonumber \\
&&\left.-(1+R)m_2\cos(2\alpha_2-\delta)-
Rm_3\cos\delta\right],
\label{eq6}\\
\dsol&=& F_{\tau}c_{12}\sin 2\theta_{12}s_{23}^2\left[m_1^2+m_2^2
+2m_1m_2\cos(2\alpha_2-2\alpha_1)\right]/\left[2(m_2^2-m_1^2)\right]
,\label{eq7}\\
{d\delta}\over {dt}&=&-\frac{F_{\tau}m_3\sin 2\theta_{12}\sin
2\theta_{23}}{2\theta_{13}(m_3^2-m_1^2)}
\times\left[m_1\sin(2\alpha_1-\delta)
-(1+R)m_2\sin(2\alpha_2-\delta)+R m_3\sin\delta\right]\nonumber  \\
&&-2F_{\tau}\left[\frac{m_1m_2s_{23}^2\sin(2\alpha_1-2\alpha_2)}{(m_2^2-m_1^2)}
+m_3s_{12}^2\left(\frac{m_1\cos 2\theta_{23}\sin 2\alpha_1}{(m_3^2-m_1^2)}
+\frac{m_2c_{23}^2\sin (2\delta-2\alpha_2)}{(m_3^2-m_2^2)}\right)\right.
\nonumber  \\
&&\left.+m_3c_{12}^2\left(\frac{m_1c_{23}^2\sin (2\delta-2\alpha_1)}
{(m_3^2-m_1^2)}
+\frac{m_2\cos 2\theta_{23}\sin 2\alpha_2}{(m_3^2-m_2^2)}\right)\right]
, \label{eq8}\\
 {d\alpha_1}\over {dt} &=&-2F_{\tau}\left [m_3\cos 2\theta_{23}
\frac{(m_1s_{12}^2\sin 2\alpha_1+(1+R)
m_2c_{12}^2\sin 2\alpha_2)}{(m_3^2-m_1^2)}\right. \nonumber \\
&&\left.+\frac{m_1m_2c_{12}^2s_{23}^2\sin(2\alpha_1-2\alpha_2)}
{(m_2^2-m_1^2)}\right],\label{eq9}\\
 {d\alpha_2}\over {dt} &=&-2F_{\tau}\left [m_3\cos 2\theta_{23}
\frac{(m_1s_{12}^2\sin 2\alpha_1+(1+R)m_2c_{12}^2\sin 2\alpha_2)}
{(m_3^2-m_1^2)}\right. \nonumber \\
&&\left.+\frac{m_1m_2s_{12}^2s_{23}^2\sin(2\alpha_1-2\alpha_2)}{(m_2^2-m_1^2)}
\right] \label{eq10}
\ea
\par
\noindent

where $R=(m_2^2-m_1^2)/(m_3^2-m_2^2)$, $P_1=s_{12}^2s_{23}^2$, $P_2 = c_{12}^2s_{23}^2$,
$P_3 = c_{13}^2c_{23}^2$, $G_3 = 0$, but
\ba
G_1&=&-\frac{1}{2}s_{13}\sin 2\theta_{12}\sin 2\theta_{23}\cos \delta
+ s_{13}^2c_{12}^2c_{23}^2, \nonumber\\
G_2&=&\frac{1}{2}s_{13}\sin 2\theta_{12}\sin 2\theta_{23}\cos \delta
+ s_{13}^2s_{12}^2c_{23}^2  \label{eq11}
\ea
\par
\noindent

In the case of MSSM with  $\mu \ge M_{\rm S}$,
\par
\noindent
\ba F_{\tau}&=&{-h_\tau^2}/{\left(16\pi^2\cos^2\beta\right)},\nonumber\\
F_u&=&\left(1\over{16\pi^2}\right)\left({6\over5}g_1^2+6g_2^2-
6{h_t^2\over\sin^2\beta}\right),\label{eq12}\ea
\par
\noindent
but, for $\mu \le M_{\rm S}$,
\par
\noindent
\ba F_{\tau}&=&{3h_\tau^2}/\left( 32\pi^2\right),\nonumber \\
F_u&=&\left(3g_2^2-2\lambda-6h_t^2 \right)/\left(16\pi^2\right)
\label{eq13}\ea

We would also need  the CP-violation parameter 
defined through Jarlskog invariant \cite{jarlskog}
\be  J_{CP} ~~ = ~~ \frac{1}{8}\sin 2 \theta_{12}\sin 2 \theta_{23}
\sin 2 \theta_{13} \cos \theta_{13}  \sin\delta \label{eq14}
\ee

Since in the case of nonzero Majorana phases, the new phenomenon
of damping of mixing angle magnification occurs, we first discuss this
case, followed by the threshold corrections specific to this case
before presenting the new results for the low energy predictions
in both cases.

\section{III~Magnification Damping by Majorana Phases and Input Parameters}

 In this section we discuss  damping of 
radiative magnification caused by
Majorana phases. While providing guidelines for the choice of
input parameters we also point out some general features of the
solutions in the context of the high scale mixing unification
model. Throughout this paper all quantities with superscript zero
indicate input parameters at the 
GUT-seesaw scale $M_R = 10^{15}$ GeV.

For our calculation, we use the low energy data for the CKM matrix
given by PDG \cite{pdg} and we take into account the appropriate
renormalization corrections to obtain their values at the
GUT-seesaw scale. Due to the dominance of the top quark Yukawa
coupling the one-loop renormalization corrections give \cite{falk},
\ba {|V^0_{ub}|\over |V_{ub}|} &=& {|V^0_{cb}|\over  |V_{cb}|} \nonumber\\
 &\simeq& exp\left[ - {y_{top}^2\over 16\pi^2} \ln{M_R\over M_Z}\right]
\simeq 0.83 \label{eq15}\ea \noindent while all other elements are
almost unaffected. Using \cite{pdg}, \ba
 |V_{ub}| &=& 0.0037, |V_{cb}|=0.0413, \nonumber\\
 |V_{us}| &=& 0.2243,  \delta = 60^{\circ}\pm {14}^{\circ}, \label{eq16}
\ea
\noindent
 corresponding to
\be  J_{CP}^{CKM} = 2.89 \times 10^{-5} \label{eq17}\ee the input
values for the  CKM matrix at the GUT-seesaw scale are, \be  \sc^0
= 0.0031, ~~\sa^0 = 0.034, ~~\sin\theta_{12}^0 = 0.224, ~~\d^0 = 60^{\circ},
\label{eq18}\ee \noindent which yield \be  J_{CP}^0 = 2 \times
10^{-5} \label{eq19}\ee \noindent where we have fixed $~\d^0 $ and
$J_{CP}^0$  at its  central value after
 CKM extrapolation.

One point of utmost importance in this analysis is  the nature of
tuning needed in the neutrino mass eigenvalues  which are inputs
at the GUT seesaw scale.  For quasi-degenerate neutrino masses
having a common
 mass $m_0$ ,
\be (m_2 - m_1) = {\Dto\over 2m_0}, ~~(m_3-m_2) = {\Dtht\over 2m_0}
\label{eq20}\ee
Using the experimental data from solar and atmospheric neutrino oscillations,
$\Dto = 8\times 10^{-5}$ eV$^2$ and $\Dtht = 2.4\times 10^{-3}$ eV$^2$
yields,
\ba
(m_2 - m_1)(eV) &=&0.004, ~~0.0002 , ~~0.0001,~~0.00004 \nonumber\\
(m_3-  m_2)(eV) &=& 0.12 , ~~0.006 , ~~0.003, ~~0.0012  \nonumber \\
  m_0(eV)       &=& 0.01, ~~0.2   , ~~0.4   , ~~1.0   \label{eq21}
\ea This clearly suggests that if we confine ourselves to values
of neutrino masses to be accessible to ongoing beta decay and
double beta decay processes i.e. $m_0 = 0.1 - 1.0$ eV, fitting the
experimental  data  on $\Delta m_{\odot}^2$   requires  tuning
between $m_1$ and $m_2$ at least up to fourth place of decimals
while fitting the data on $\Delta m_{atm}^2$ needs tuning at least
up to the third place of
 decimals between $m_2$ and $m_3$. The region of small  common mass where the tuning improves by 1-2 orders
is inaccessible to these experiments and also the RGEs do not
succeed in achieving the desired magnification when $m_0 << 0.1$
eV. We again emphasize the point noted in \cite{mpr1} that
successful radiative magnification requires that $m_0\geq 0.1$ eV.
Our idea can therefore be ruled out if there is no evidence for
Majorana neutrino mass in the next round of neutrino double beta
decay searches \cite{bb} as well as, of course, by the measurement of
$\theta_{13}$.

It is  known that RGEs of neutrino mixing angles exhibit  quasi-fixed points 
at low energies  and these predicted quasi-fixed points are not
in agreement with the observed experimental data \cite{casas}.  
Exploiting the  HUM  hypothesis we have shown \cite{mpr1,
  mpr2,mpr3} that these quasi-fixed ponts can be avoided in supersymmetry 
throgh radiative magnification which is achieved  
 by appropiate fine-tuning of  quasi-degenerate neutrino mass eigen values and 
for larger values of $\tan\beta$. Such type of fine-tuning has been discussed
in ref.\cite{mpr1}.

A useful general property of the RGEs with quasi-degenerate
neutrinos is the following scaling behavior: for a set of
solutions with  masses $m_i$, mixings, phases, and mass squared
differences $\Delta m_{ij}^2$, any other  set of masses scaled to
$\kappa m_i$ are also solutions with the same mixings and phases
but with mass squared differences scaled to $\kappa^2 \Delta
m_{ij}^2$ where $\kappa$ is the common scaling factor for all mass
eigenvalues. This property serves as a useful tool to
 derive other allowed solutions from one set of judiciously chosen numerical solutions to the RGEs.

  From the structure of eq.(5)- eq.(10) it is clear that for a set
of solutions $\ao, \at,$ and $\d$ , there exists another set of
solutions $-\ao, -\at,$ and $-\d$ . The initial choice of input
Majorana phases $(\ao^0, \at^0) = (0, 0), (\pi/2 , \pi/2 ), (\pi,
\pi)$ continue to retain their values to low energies up to a good
approximation.

As noted in ref. \cite{mpr1} in the top down approach the RGEs for
mass eigenvalues given in eq.(4) not only decrease the masses but
they also decrease their differences such that at low energies
$m_i \sim m_j$.
The radiative magnification of the three mixing angles takes place for
quasi-degenerate neutrinos because of smallness of the mass differences
$m_i-m_j(i\neq j = 1,2,3)$ that occurs ~in the denominators of
the RHS of RGEs for the mixing angles in eq.(5)- eq.(7). For the same reason
large changes are also expected in the phases.
During the course of RG evolution in the top-down approach $m_i \to m_j$ and
the rate of magnification for mixing angles increases. Equivalently the solar and
atmospheric mass differences decrease from their values at $\mu = M_R$ to
approach  their experimental
values at $\mu = M_Z$ and the radiative magnification occurs for all the
three mixing angles. While $\sin\theta_{12}$ and $\sa$  attain their respective large values
$\sc$ remains small at low energies in spite of its magnification
because its high scale
starting value derived from the quark sector is much smaller($= |V^0_{ub}|$).

Note that since the input values for $\sin\theta_{12}^0$ , $\sa^0$, and
$\sc^0$ are small with much smaller values for
$\sin^2{\theta^0_{ij}}$ , but $\cos^2{\theta_{ij}^0} \simeq
1(i\neq j = 1, 2, 3)$, it is clear from the RHS of eq.(5) that
initially substantial contribution to the rate of magnification of
$\sa$ is caused by the dominant $c_{12}^2$ term, but $s_{12}^2$
term contributes only later during the course of evolution.

   The following considerations hold when both the Majorana phases are small.
In eq.(7) the coefficient on the RHS is proportional to $ s_{12}s_{23}^2$
whose initial value is very small. Therefore the magnification of 
$\sin\theta_{12}$
requires very  small difference between $m_2^0$ and
$m_1^0$ compared to the difference
$m_3^0-m_2^0$. On the other hand the dominant term in the RHS of
eq.(5) is proportional to $s_{23}c_{23}^2c_{12}^2$ which is much
larger near $\mu = M_R$ than the corresponding coefficient in
eq.(7). The basic reason for this is essentially the smallness of
CKM mixings near the GUT-seesaw scale which
is totally different from the situations where initially there are
large mixings \cite{antusch1, xing}.~Therefore in the HUM model radiative
magnification of $\sa$
can occur for larger mass difference between $m_3^0$ and $m_2^0$ compared to
$m_2^0- m_1^0$ which is required for the magnification of $\sin\theta_{12}$ 
through eq.(7).

   In the presence of finite  values of Majorana phases the situation
undergoes a quantitative change even if the smallness conditions
in the initial mass differences are satisfied. In general damping
in the magnification of the mixing angles $\theta_{12}$ and
$\theta_{23}$ sets in whenever $\cos 2\alpha_i$ ($i = 1, 2$), or
$\cos 2(\ao - \at)$  occurring in eq.(5) and eq.(7) deviate from
$+1$.

To see the reason for magnification damping  in a qualitative
manner, note that the presence of Majorana phases is equivalent to
multiplying the mass of the neutrino by that phase (i.e.
$m_0e^{i\alpha}$). Now recall the well known case where for $\ao
-\at = (2n +1)\pi/2$ corresponding to the case of opposite
CP-Parity of $\nu_1$ and $\nu_2$, magnification of $\theta_{12}$
is prevented. Similarly if $\ao = \at = \pi/2$, the CP-Parity of
$\nu_3$ is opposite to that of $\nu_1$ or $\nu_2$ and the mixing
angle $\theta_{23}$ can not be magnified. In general values of
$|\ao|, |\at|$ different from $0$ or $\pi$ cause damping to the
magnification of mixings, the damping being stronger for negative
values of cosine functions occurring in the RHS of eq.(5) and
eq.(7) . But as we find in spite of such dampings radiative
magnification of the mixing angles can still be realized if these
cosines have positive values $< 1$ at lower scales.  It is clear
from eq.(5) that  if $\alpha_2 = \alpha_1 = \pi/2$ the
magnification of $\sa$ can not occur even though the smallness
condition on the mass difference  $m_3^0 - m_2^0$ is satisfied. If
$\alpha_2 \simeq \pi/2$ this magnification is badly affected. In
fact even if $\ao^0$ is small, but if $\at > {\pi\over 4}$ the
dominant magnifying term in eq.(5) would start decreasing and it
would need still smaller value of the initial mass difference
$m_3^0 - m_2^0$ to achieve magnification. As a result this region
of Majorana phases would be restricted by the low energy data on
$\Dtht$ since the latter is proportional to the mass difference
$(m_3 - m_2)$.

   From eq.(7) we see that no magnification can occur for
$\sin\theta_{12}$ if $|\alpha_2 - \alpha_1|
\simeq \pi/2$ which  can be satisfied for
$\alpha_2 \simeq 0$, $\alpha_1 \simeq \pi/2$ or for $ \alpha_1 \simeq 0$,
$\alpha_2 \simeq \pi/2$.   Eq.(5) and eq.(7) permit magnifications of $\sa$ and
$\sin\theta_{12}$ for small values of the Majorana phases $\ao, \at < {\pi\over 4}$ .
Also when $\at$ is small, magnification is allowed for all values
of $\ao = 0 - \pi$ except in the region
$\ao \simeq {\pi \over 2}$
where the magnification of the solar mixing angle is strongly
damped out for similar choice of the mass difference
$m_2^0-m_1^0$ for which magnification takes place
in other cases.

       When $\alpha_1 \simeq \alpha_2 \simeq 0 $, it is clear
from eq.(8) that ${d\delta
\over dt}$ is directly proportional to $\sin\delta$  and the RG solution for the
leptonic Dirac phase approaches $\delta \to 0$ as $\mu \to M_Z$.
A general consequence of the RGEs for the Majorana phases is that if
we start with the initial condition $\alpha_1^0 = \alpha_2^0$
at the GUT-seesaw scale, they would
obtain equal values during the course of evolution \cite{antusch2}.
Therefore if we start with vanishing Majorana phases at the GUT-seesaw scale
they continue to remain zero at low energies.

    Even though certain input values of Majorana
phases might appear to be affecting
magnification of mixings at high scales, their role to cause damping becomes
different at  lower scales as the supplied phases  change by RG evolution.
     Particularly the damping becomes most severe
for the choice $\ao^0 \simeq \at^0 = {\pi/2}$
since these phases which damp out magnification of both the mixing angles
to start with remain almost unaltered during the course of evolution.

     Although certain choices of  input values of Majorana phases  can
cause partial
or total  damping of magnification of mixing angles, moderate
damping with other choices can be utilised to obtain smaller mass squared differences
at low energies. Exploiting this mechanism we show in Sec.VI how we
identify a region in parameter space  where $\Dto$ needs no
threshold corrections to be in agreement with the experimental
data. On the other hand with MSSM spectrum below the GUT-seesaw
scale, as the super-partner masses do not appear to be very close
to $\mu = M_Z$, small threshold corrections to mass squared
differences due to super-partners are quite natural in any such
model containing the MSSM. In the next section we present
derivation of new formulas showing how Majorana phases also affect
threshold corrections in all models having quasi-degenerate
neutrino masses.

\section{IV~Threshold Effects with Majorana Phases}

With MSSM as the effective theory below the GUT-seesaw scale and
 non-degenerate super-partner masses new
threshold corrections at $\mu = M_Z$ arise in any model having
quasi-degenerate neutrino masses \cite{chun}. They have been
derived ~in the quasi-degenerate case in the limit $\theta_{13}
\to 0$ and $\theta_{23} \to \pi/4$ but without phases \cite{mpr2}.
Following the same procedure and noting that the mixing matrix
elements $U_{\alpha i},(i= 1, 2)$ now contain Majorana phases,  we
obtain the new generalized analytic formulas which are valid in
all models with quasi-degenerate spectrum,

\ba (\Delta m_{21}^2)_{th}
&=&4\rm m^2\left[\left(s_{12}^2\cos 2\at- c_{12}^2\cos 2\ao\right)T_e + \left(
c_{12}^2\cos 2\at- s_{12}^2\cos 2\ao\right){(T_\mu + T_\tau)\over 2}\right],
\nonumber \\
(\Delta m_{31}^2)_{th}
&=&4\rm m^2\left[ -T_e c_{12}^2\cos 2\ao+ \left(1 - s_{12}^2\cos 2\ao\right)
{(T_\mu+T_\tau)\over 2}\right], \nonumber \\
(\Delta m_{32}^2)_{th} &=&4\rm m^2\left[ -T_e s_{12}^2\cos 2\at+
\left(1 - c_{12}^2\cos 2\at\right) {(T_\mu+T_\tau)\over 2}\right]
\label{eq22} \ea Here $m$ represents the common mass of
quasi-degenerate neutrinos. The functions $T_{\alpha}, (\alpha =
e, \mu, \tau)$ are one loop factors obtained by evaluation of
corresponding Feynman  diagrams, especially with wino and charged
slepton exchanges \cite{chun}.

\ba
T_{\alpha}=(g^2/ {32\pi^2})[(x_{\mu}^2-x_{\alpha}^2)/
(y_{\mu}y_{\alpha})+((y_{\alpha}^2-1)/ y_{\alpha}^2)
ln(x_{\alpha}^2)- ((y_{\mu}^2-1)/ y_{\mu}^2)ln(x_{\mu}^2)]\label{eq23}
\ea
\par
\noindent
where $y_{\alpha} = 1-x_{\alpha}^2$, $x_{\alpha} = M_{\alpha}/M_{\tilde w}$,
$M_{\alpha} =$ charged slepton mass, and $M_{\tilde w} =$  wino mass.
and the loop-factor has been defined to give $T_{\mu} = 0$ without any loss of
generality.
Depending upon the allowed low energy values of the Majorana phases
 the threshold corrections may vary for given values of the super-partner masses.
In particular when $\ao = \at = 0, n\pi$, these generalized
formulas reduce to those obtained in ref. \cite{mpr2},

\ba
(\Delta m_{21}^2)_{th}
&=&4\rm m^2\cos^2\theta_{12}[-T_e+(T_\mu+T_\tau)/2], \nonumber \\
(\Delta m_{32}^2)_{th}
&=&4\rm m^2\sin^2\theta_{12}[-T_e+(T_\mu+T_\tau)/2], \nonumber \\
(\Delta m_{31}^2)_{th}
&=&4\rm m^2\cos^2\theta_{12}[-T_e+(T_\mu+T_\tau)/2] \label{eq24}
\ea
\par
\noindent

Including threshold corrections along with the RG-evolution effects the mass squared differences are evaluated using
\par
\noindent
\be\Delta m_{ij}^2=(\Delta m_{ij}^2)_{\rm RG}+
(\Delta m_{ij}^2)_{\rm th} \label{eq25}\ee
\par
\noindent
where the RG-evolution effects from $M_{\rm R}$
to $M_{\rm Z}$ are computed using the procedure discussed in
Sec.II - Sec.III. For several allowed mass ratios,
$M_{\tilde {\rm e}}/M_{{\tilde \mu},{\tilde \tau}} \simeq 1.4-2.8$,
consistent with
an  inverted hierarchy in the charged-slepton sector
we find that the RG and the threshold corrections together are in good
agreement with
the available experimental data as discussed in Sec.V and in Table I.
Small changes in the mixing angles due to  threshold corrections in the
masses are easily compensated by very small changes in the input
  mass eigenvalues leading to predictions given 
in Table.I in the next section.
In Sec.VI we also present solutions without any threshold corrections.

\section{V~Low Energy Predictions in the Leptonic Sector}

In this section we present  the predictions of our model as
solutions of RGEs with Dirac and Majorana phases. As explained in
ref. \cite{mpr1} at first we follow bottom up approach to
obtain information on the Yukawa and gauge couplings at the
GUT-seesaw scale to serve as inputs in the top down approach.
The extrapolated values of the gauge and the Yukawa couplings for
$\tan \beta = 55 $
at $M_R = 10^{15}$ GeV
are:
$g_1^0 =0.6683, g_2^0 = 0.6964,  g_3^0 = 0.7247 , h_{\rm {top}}^0 = 0.8186,
h_b^0 = 0.6437$, and $h_{\tau}^0 = 0.7105 $.
We choose
the high seesaw scale  $M_R = 10^{15}$ GeV where the mixing angles and
the Dirac phase are those given in eq.(18).
We use high scale input mass eigenvalues
as discussed above.
Using solutions of RGEs combined with such threshold 
corrections discussed in
Sec.IV in
the appropriate cases we treat the results as acceptable if they are within
$4\sigma$ limit of the available data from neutrino oscillation experiments
although a substantial part  of our solutions in the parameter space
are either in the best fit region or within $2\sigma$ to $3\sigma$
limits \cite{maltoni},
\ba
\Dto   &=& (6.8 - 9.3)\times 10^{-5} eV^2, \nonumber\\
\Dtho  &=& (1.8 - 3.5)\times 10^{-3} eV^2, \nonumber \\
\sin^2\theta_{12} &=& 0.22 - 0.44, \nonumber\\
\sin^2\theta_{23} &=& 0.31 - 0.71, \nonumber\\
\sin^2\theta_{13} &\le& 0.058 \label{eq26}
\ea

Since the other factor occurring in $J_{CP}$ is almost determined
from the solar and atmospheric mixing angles, $(\sin 2\theta_{12}
\sin 2\theta_{23})/ 8  \simeq 0.1 $ the model dependent quantity
$|\rm {Im} U_{e3}| \simeq O(0.1)$
 if the prediction is to be verified by long base line neutrino oscillation  experiments in near future.
The predictions of $\theta_{13}$ and  $\d$ vary from model to model.
In the present  case the initial value of $V^0_{ub}$ including the Dirac phase
$\d^0$ and the input values of the Majorana phases determine $\theta_{13}$
and $\d$  at low energies by RG evolution while matching
the experimental values of the other two neutrino mixing angles by
radiative magnification. We find that in the presence of Majorana phases our model predicts values of
$|\rm {Im} U_{e3}|$ contributing to $|J_{CP}|
\simeq O(10^{-2})$ accessible for measurement by neutrino oscillation
 experiments.

As discussed in Sec.III the most flexible choice
in the parameter space for which radiative magnification
of mixings takes place easily is with
$\at^0 < 45^0$  or $\at^0 > 135^0$  and varying $\ao^0$.
We present below details of
results obtained on mass squared differences, mixing angles, low
energy values of different phases and
the CP-violation parameter. In all cases
the Dirac phase of the PMNS matrix
has been fixed at $60^{\circ}$ at the GUT-seesaw scale. The other class of
solutions obtained using the magnification damping criteria and which needs no threshold corrections due to 
super-partner masses will be presented in Sec.VI.

\subsection{V.1~Low Energy Predictions for $\theta_{13}$ and  Majorana Phases}

Since  Majorana phases do not occur in eq.(4) and Dirac phase
dependent terms are very small, the evolution of mass eigen values
are almost the same as reported earlier \cite{mpr1}. They decrease
from their high scale input values  and tend to converge towards
one another narrowing their differences until agreement with
experimental data on $\Dtho$ and $\Dto$ are obtained with or
without small threshold corrections. Even in the presence of
phases the radiative magnification to bi- large  neutrino mixings
maintain the similar correlations as before: Decreasing
differences between the mass eigenvalues is accompanied by
increasing magnification of mixing angles and finally their
largest predicted values are obtained at $\mu = M_Z - M_S$ where
$M_S = 1$  TeV is the SUSY scale. For example with the input value
of one Majorana phase set to a vanishing value at the GUT-seesaw
scale ($\at^0 = 0.0$), the radiative magnification to bi-large
neutrino mixings takes place for all values of the other Majorana
phase $\ao^0 = 0 - \pi$  except at $\ao^0 \simeq {\pi\over 2}$
where almost total damping occurs for the  magnification of $\sin\theta_{12}$.

Fig. 1(a) shows  predicted variation of $\sc$ at low energies  as a function of
 $\ao^0$. It is clear that starting from $\theta_{13} = 3.5^{\circ}$ at
$\ao^0 = 0 $ the value of $\theta_{13}$ continuously increases
reaching a maximal value at $\ao^0 \simeq 85^{\circ}$. Again after
the disfavored region at $\ao^0\simeq 90^{\circ} \pm 5^{\circ}$,
it decreases from its maximal value continuously in the second
quadrant till it reaches  low  value at $\ao^0 \simeq \pi$. The
predicted range of $\theta_{13} \simeq 3.5^{\circ} - 10^{\circ}$
and the maximum predicted value is nearly $25\%$ smaller than the
current $4\sigma$ upper bound. This prediction is new and covers a
wider range compared to that found in ref. \cite{mpr1} where the
PMNS matrix had no phase.

The low energy values of Majorana phases corresponding to
the input values of $\ao^0$ are shown in Fig.1(b)and Fig.2(a).
We note from Fig. 1(b) that corresponding to input $\at^0 = 0.0$
the Majorana phase $\ao(M_Z)$
at low energies increases from $0^{\circ}$
to $20^{\circ}$ when the input $\ao^0$  varies from  $0^{\circ}$ to
$ 85^{\circ}$  in the first quadrant but $\ao(\Z)$  varies from
$155^{\circ}$ to
$180^{\circ}$  when $\ao^0$ varies in the second quadrant.
 From Fig. 2(a) we find  that though the value of $\at^0 = 0$ at
the GUT-seesaw
scale, its value at low energies is
non-vanishing with $\at(M_Z)
\simeq  0^{\circ}$  to $-8^{\circ}$ when $\ao^0$ varies over  the first
quadrant, but  $\at(M_Z)= 8^{\circ}$ to $0^{\circ}$ when
$\ao^0$ is in the second quadrant.

\subsection{V.2~Predictions for Dirac Phase and Leptonic CP Violation}

Two  interesting features of the model  are  the unification of the CKM
Dirac phase with the PMNS Dirac phase and also the unification of
quark CP-violation with the leptonic CP-violation at the GUT-seesaw scale.
We summarize low energy predictions of leptonic CP-violation in two different
cases: (i) When Majorana phases are absent in the PMNS matrix at high scales,
(ii) When the high scale PMNS matrix contains Majorana
phases as unknown parameters.\\

\par\noindent{\bf V.2.A ~CP-Violation without Majorana Phases}\\

When the two Majorana phases are set to zero at the GUT see saw scale,
the PMNS matrix unifies with
the CKM matrix including its Dirac  phase.
As a consequence of the RGEs in the top down approach in the $\theta_{13} \to 0$
limit, the Dirac phase rapidly decreases to approach its trivial fixed point
$\delta \to 0$. Both the Majorana phases also continue to maintain their
vanishing values in this limiting case at all energies below the GUT-seesaw scale and  we obtain
\ba
\delta &\simeq& 0.3^{\circ} - 1^{\circ}, ~~\theta_{13} \simeq 3.5^{\circ}
 - 4.5^{\circ},
 \nonumber\\
 J_{CP} &\simeq&  (5 - 7.6)\times10^{-5} \label{eq27}
\ea
Thus, without initial Majorana phases, the CP-violation parameter is a few 
times larger than
the corresponding low energy value in the CKM matrix even though
the low energy value of the Dirac phase is suppressed by  a factor $
\simeq 10^{-2}$  than the corresponding CKM Dirac phase. This occurs due to the predicted value of the
 CHOOZ angle which is nearly $20$ times larger than its counterpart in
the quark sector ($|V_{ub}|$) at
low energies.\\

\par\noindent{\bf V.2.B  ~CP-Violation  with Majorana Phases}\\

The predictions on the leptonic Dirac phase and CP-violation parameter
change substantially at low energies in the presence of finite input values
of Majorana phase(s) at the GUT-seesaw scale.
The variation of predicted leptonic Dirac phase
$\d$ and the leptonic CP-violation parameter
$J_{CP}$ at low energies  with the input Majorana phase
$\ao^0$ are shown in Fig. 2(b) and Fig. 3(a), respectively.
It is clear from
Fig. 2(b) that
the leptonic Dirac phase varies from $0.3^0$  to $-70^{\circ}$ when
$\ao^0$ is in the first quadrant, then it becomes
positive and varies from $80^{\circ}$ to $0.3^{\circ}$
when $\ao^0$ is in the second quadrant.
The variation of the leptonic
CP-violation parameter is correlated with the variation of the
leptonic Dirac phase with $\ao^0$. In Fig. 3(a) the low-energy values
of $J_{CP}(M_Z)$
in the first(second) quadrant  are negative(positive). It is clear from Fig.
 3(a) that $J_{CP}(M_Z)$ varies from
$ -7\times 10^{-5}$ to $-0.03$ when $\ao^0$ is in the first quadrant
, but then it changes sign and varies from $ 0.038$ to
$7\times 10^{-5}$ when $\ao^0$ is in the second quadrant.

The evolution of Dirac and Majorana phases are shown in Fig. 3(b)
corresponding to the input values $\ao^0 = 30^{\circ}, \at^0 = 0.0$. While
$\ao(\mu)$ is found to decrease by nearly $50 \%$, a nonzero value for
$\at(\mu)$  is found to have been generated at $\mu = \Z$ from
its vanishing boundary value at the GUT-seesaw scale.
The solid(dotted) line in Fig. 3(b)
represents the evolution of the leptonic(CKM) Dirac  phase with unification
at the GUT-seesaw scale. While the CKM Dirac phase remains constant the
leptonic Dirac phase is found to evolve to nearly $-35^{\circ}$ at low energies starting from its unified value of $60^{\circ}$ at the GUT-seesaw scale.

One of the most interesting and novel outcome of this analysis
is the prediction of the leptonic CP-violation starting from the corresponding
baryonic CP-violation as embodied in the CKM matrix as shown in Fig. 4(a) and 
Fig. 4(b)
where the dotted line represents very slow evolution of the CKM CP-violation
parameter. While in Fig. 4(a) we present positive values of the parameter
in Fig. 4(b) we present negative values depending upon the initial choice of 
Majorana phases.
The dot-dashed line in
Fig. 4(a) represents the evolution of the leptonic CP-violation parameter
starting from its unified CKM value of ~$2\times 10^{-5}$.
In the absence of any
Majorana phase  the radiative magnification caused due to quasi-degenerate
neutrinos has enhanced it by nearly a factor of only $\simeq 3.5 $ making the
low energy prediction $J_{CP} = 7.6\times 10^{-5}$.

     In the presence of non-vanishing Majorana phase(s), the parameter
evolves to much larger values at low energies
as shown by the dashed and solid lines in Fig. 4(a) corresponding to 
$J_{CP}(M_Z) \simeq 0.01 - 0.03$. 
~In Fig. 4(b) the negative values of the parameter at low energies are in the 
range  $J_{CP}(M_Z) \simeq -0.013$ to $-0.036$. It is clear from these two
figures that major part 
 of magnification of $|J_{CP}|$ occurs in the region 
$\mu = 10^{5}$ GeV to $10^{3}$ GeV.
We note that the continued smallness of the CP-violation parameter at high
scales for $\mu = 10^7$ GeV to $10^{15}$ GeV might have some cosmological significance.   
Although these predicted values of the re-phasing invariant parameter, $|J_{CP}(M_Z)| \simeq 0.02 - 0.04$ 
at low energies that relates only the Dirac phase, are 
accessible for measurement by long baseline neutrino oscillation experiments,
 we have also
estimated the model predictions on the two other invariants relating to 
Majorana phases \cite{jarlskog, picariello} for the sake of completeness,
\ba
 S_1 &=& Im\left( U_{\nu_e\nu_1}U_{\nu_e\nu_3}^*\right) = \frac{1}{2}
\sin 2\theta_{13}\cos\theta_{12}\sin (\ao+\delta) , \nonumber\\
S_2 &=& Im\left( U_{\nu_e\nu_2}U_{\nu_e\nu_3}^*\right) = \frac{1}{2}
\sin 2\theta_{13}\sin\theta_{12}\sin (\at+\delta)  \nonumber
\ea  

The predicted values of these two parameters are shown in Fig. 5(a)
as a function of the input Majorana phase $\ao^0$ while $\at^0 = 0$. We find
that starting from   $S_1 \simeq S_2 \simeq 10^{-4}$ these parameters 
are bounded by 
 $|S_1| \le 0.1$ and $|S_2| \le 0.14$  for this choice of Majorana phases. 
If in future, experiments 
on neutrinoless double beta decay determine one of the two Majorana phases,
then with possible information on $\theta_{13}$ and the leptonic Dirac phase 
from neutrino oscillation experiments, the predictions on one of these parameters might be tested.   

\subsection{ V.3~Neutrinoless Double Beta Decay, Tritium Beta Decay and Cosmological Bound}

The search for neutrinoless-double beta decay by Heidelberg-Moscow
experiment has obtained the upper limit,
\be
  <m_{ee}> < ( 0.33 - 1.35 ) ~{\rm {eV}} \label{eq28}\\
\ee
which overlaps the range that will be covered in the planned experiments
 \cite{cremo}.
Similarly the current upper bound on the kinematical mass from
Tritium beta decay is $<m_e> < 2.2$ eV .
But the KATRIN experiment is expected to probe
$<m_e> \ge 0.2$ eV and if positive it would prove that the neutrino
masses are quasi-degenerate \cite{rnm}.
Using
\ba
<m_{ee}> &=&  | m_1 c_{12}^2c_{13}^2 + m_2 s_{12}^2c_{13}^2 e^{-2i(\at-\ao)}
+m_3 s_{13}^2 e^{2i(\d + \ao)}  |,
 \nonumber\\
<m_e>    &=& \left [ m_1^2 c_{12}^2c_{13}^2 + m_2^2 s_{12}^2c_{13}^2
+m_3^2 s_{13}^2 \right]^{1/2}
\label{eq29}
\ea
we have evaluated these effective masses using our solutions for
mass eigen values, mixing angles and Majorana and Dirac phases at low energies.
Some of our solutions including small threshold corrections
estimated using the formulas given in eq.(22) - eq.(23) are presented in
Table I while others without the necessity of threshold corrections
are discussed in Sec. VI and presented in Table II. 
It is clear that not only the model predictions are in good agreement
with the current data from neutrino oscillation experiments, but also
the mass parameters are in the interesting range accessible to the Heidelberg
Moscow experiment on double beta decay 
and the KATRIN experiment on beta decay.
At this stage we emphasize that since the 
super-partner masses may be naturally
in the range of $150$ GeV $- 1$ TeV,  small threshold 
corrections in the MSSM
near the electroweak scale or even 
at $\mu = M_S = 1 $ TeV are quite natural.

        Taking into account the allowed range of positive  mass eigenvalues
this analysis predicts the sum of three quasi-degenerate neutrino
masses  to be in the range $\Sigma_i m_i = (0.7 - 1.2 )$ eV.
Recent data from WMAP gives the bound on the sum of the three neutrino masses
in~the ~range $(0.7 - 2.1)$ eV depending upon
what values one chooses for the priors.
The cosmological bound may be as large as  $3$ eV \cite{wmap, fukugita}.
In this context it is interesting to note that the
presence of Dark Energy may have strong impact on
cosmological bounds on neutrino masses taking the
lowest value of WMAP bound on the sum without priors
from $0.7$ eV to $1.4$ eV \cite{hannestad}.

We find that the values of quasi-degenerate neutrino masses
predicted in the model are consistent with the current cosmological bounds.

\section{VI~Parameter Space for No Threshold Correction}

As pointed out earlier \cite{mpr1, mpr2} and utilized in Sec. IV
and Table I the low energy solutions for $\Dto$ in the presence or
absence of Majorana phases need small threshold corrections due to
possible  spreading of sfermion masses above the electroweak scale
to bring $\Dto$ to be in agreement with the experimental data. Of
course these corrections are quite natural and must be included in
any SUSY model of quasi-degenerate neutrino masses. In the present
case a class of  our solutions for $\Dto$ need such small
threshold corrections which are readily computed as discussed
in \cite{mpr2, chun}.

In this Section we find an interesting new aspect of the radiative
magnification mechanism in the presence of Majorana phases: i.e.
there is a region in the parameter space which needs no threshold
corrections for the mass squared differences to be in agreement
with observations. In terms of initial values of the Majorana
phases we find that this region of parameter space is described to
a very good approximation by the two branches of the curve
presented in Fig. 4(b). This curve is described by the equation,

\ba
 {\rm C}os ~2(\ao^0 - \at^0 ) & = & {\rm {Const}}. \nonumber\\
\rm {Const}. & \simeq & -0.9 \label{eq30}
\ea

These solutions without threshold corrections and with Majorana
phases consistent with eq.(30) are given in Table II. It is clear
that in this case the predictions for the reactor angle
$\theta_{13}$, the Dirac phase and the CP-violating parameter
are
 accessible to long baseline experiments while the effective neutrino mass parameter is within the
range to be probed in neutrinoless double beta decay experiments
as well as the KATRIN experiment. Further the sum of the three
neutrino masses is also consistent with the cosmological bound
including those from WMAP.

It is interesting to examine the physical origin behind such solutions.
At first we note that  eq.(30) looks very much like a
strong damping condition on the radiative
magnification of $\sin\theta_{12}$ as discussed in Sec.III.
But, in reality, the values of Majorana phases change
from their high-scale  boundary values and while the
major part of  magnification takes place at lower scales,
${\rm C}os ~2(\ao(\mu) - \at(\mu))$ becomes positive and remains
in the moderate damping region.
In fact we have found from low energy values obtained from RG evolution that
${\rm C}os ~2(\ao(\Z) - \at(\Z)) = 0.49 - 0.58 $ as shown in Table. II.

Having shown that this cosine function is in the moderate
damping region it then follows from eq.(7) that the
radiative magnification of $\sin\theta_{12}$ in the presence of
such moderate damping requires smaller
mass difference between $m_2$ and
$m_1$ than the corresponding case with
weaker damping. This then leads to
solutions to RGEs with smaller $\Dto$ which needs no threshold corrections.

\section{VII~Conclusion}

In conclusion, we have extended the results of the high scale
mixing unification (HUM) hypothesis discussed in \cite{mpr1} by
including the effects of the CP phases. We have considered the
cases with and without the Majorana phases. While the mixing
unification hypothesis predicts the Dirac phase to be equal to the
CKM phase, it leaves the Majorana phases arbitrary since they have
no quark counterpart and will most likely arise from the right
handed neutrino sector. For both cases, we find consistent
quasi-degenerate neutrino mass patterns that lead to desired
amount of radiative magnification of the mixings in agreement with
data. The predictions of the model are as follows: the common mass
of the neutrinos must be larger than $0.1$ eV; the values of
$\theta_{13}$ and CP phases in the lepton sector are also
predicted at low energies. In the case without the Majorana phase,
the low energy CP violating effect is small with $J_{CP}\simeq
7.6\times 10^{-5}$, whereas for the case with Majorana phases,
$J_{CP}$ can be as large as $0.04$. An interesting phenomenon in
the latter case is the damping of radiative magnification of
neutrino mixing angles which constrains the Majorana phases to
remain in a certain range for the model to explains the two large
neutrino mixing angles. In this case, we predict a wider range of
$\theta_{13}$ from $3.5^{\circ} - 10^{\circ}$.  The larger ranges
for both $\theta_{13}$ and $J_{CP}$ are accessible to measurements
by long baseline neutrino oscillation experiments.

The quasi-degenerate neutrino masses required to achieve desired
radiative magnification, are consistent with the current
cosmological bounds including those from WMAP. They are also
accessible to the ongoing  laboratory experiments on neutrinoless
double beta decay and overlaps the range of  KATRIN experiment for
beta decay.

\begin{figure}[htbp]
  \centering
  \includegraphics[width=0.45\textwidth,height=0.3\textheight]
  {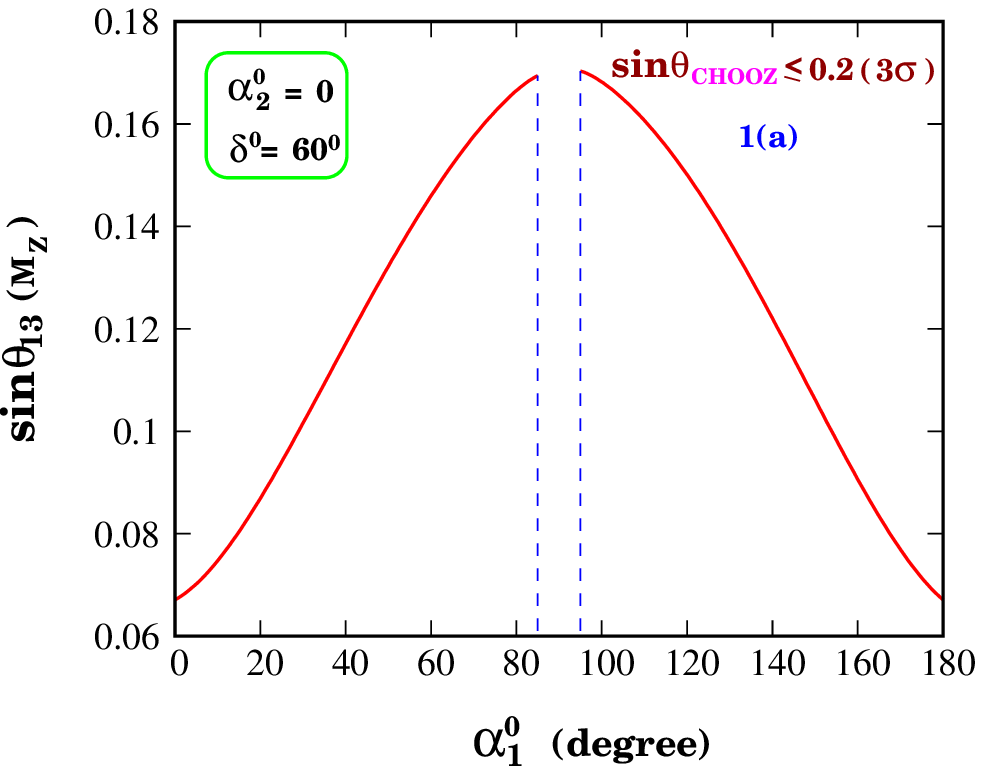}
 \includegraphics[width=0.45\textwidth,height=0.3\textheight]{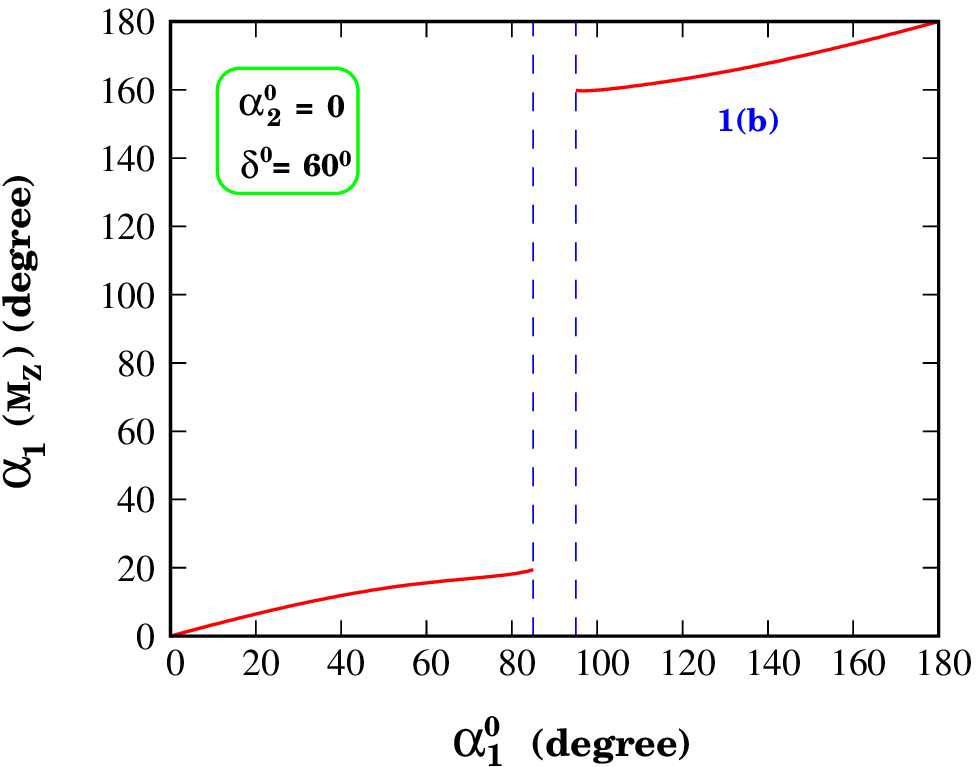}
  \caption{1(a). Prediction of CHOOZ angle as function of input Majorana phase
$\ao^0$ at the GUT-seesaw scale for $\at^0 =0$. The vertical dashed lines
define the total damping region for $\sin\theta_{12}$ magnification corresponding to
$\ao^0 = 90^{\circ} \pm 5^{\circ}$.
1(b). RGE-solution for $\ao$ at low energies as a function of the input phase at the GUT-seesaw scale.}
  \label{th13_del}
\end{figure}

\begin{figure}[htbp]
  \centering  \includegraphics[width=0.45\textwidth,height=0.3\textheight]
  {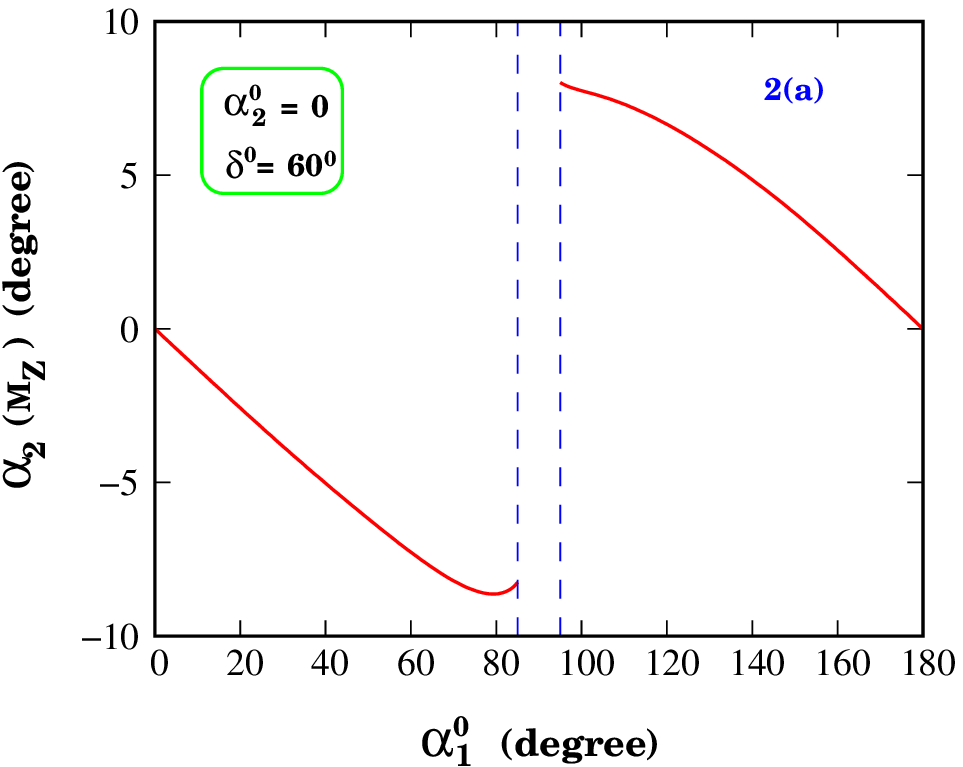}
  \includegraphics[width=0.45\textwidth,height=0.3\textheight]{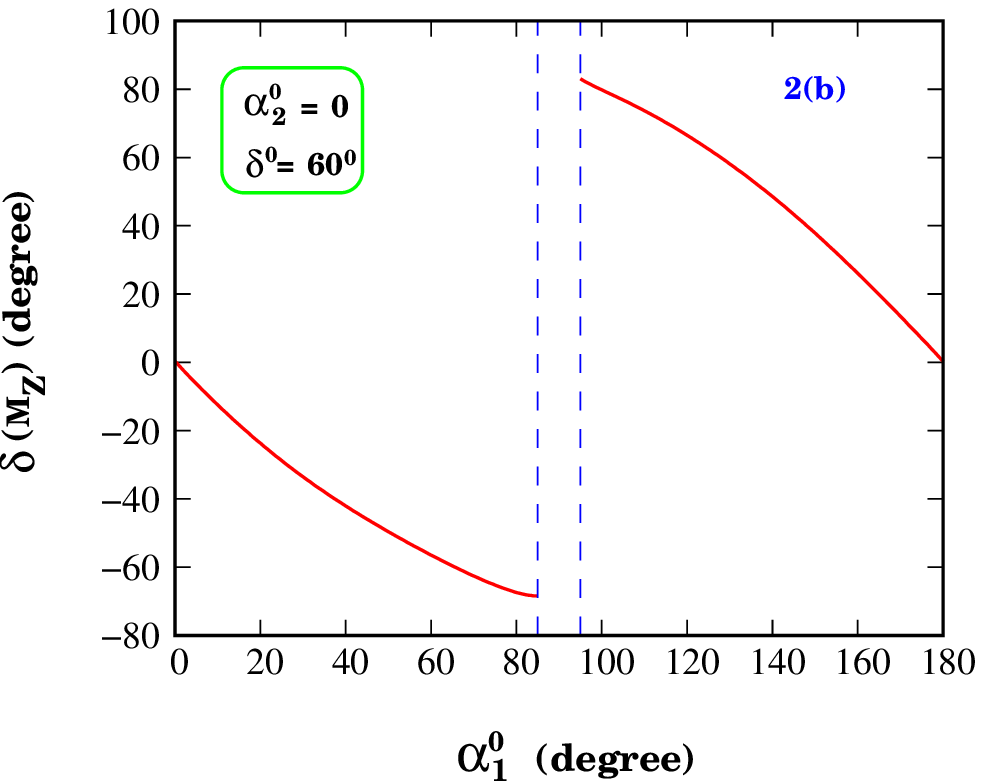}
  \caption{2(a). RGE-solution for $\at$ at low energies as a function of input $\ao^0$ at the GUT-seesaw scale.
2(b). RGE-solution for leptonic Dirac phase at low energies as a function of $\ao^0$.}
  \label{a2_del}
\end{figure}

\begin{figure}[htbp]
  \centering  \includegraphics[width=0.45\textwidth,height=0.3\textheight]
  {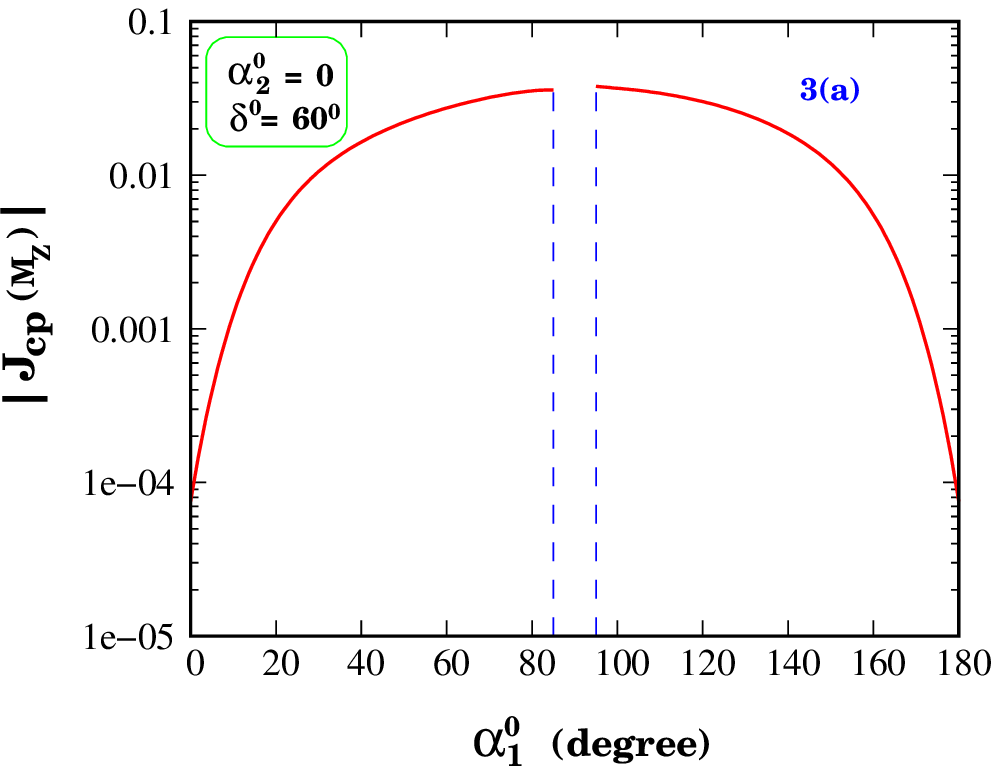}
  \includegraphics[width=0.45\textwidth,height=0.3\textheight]{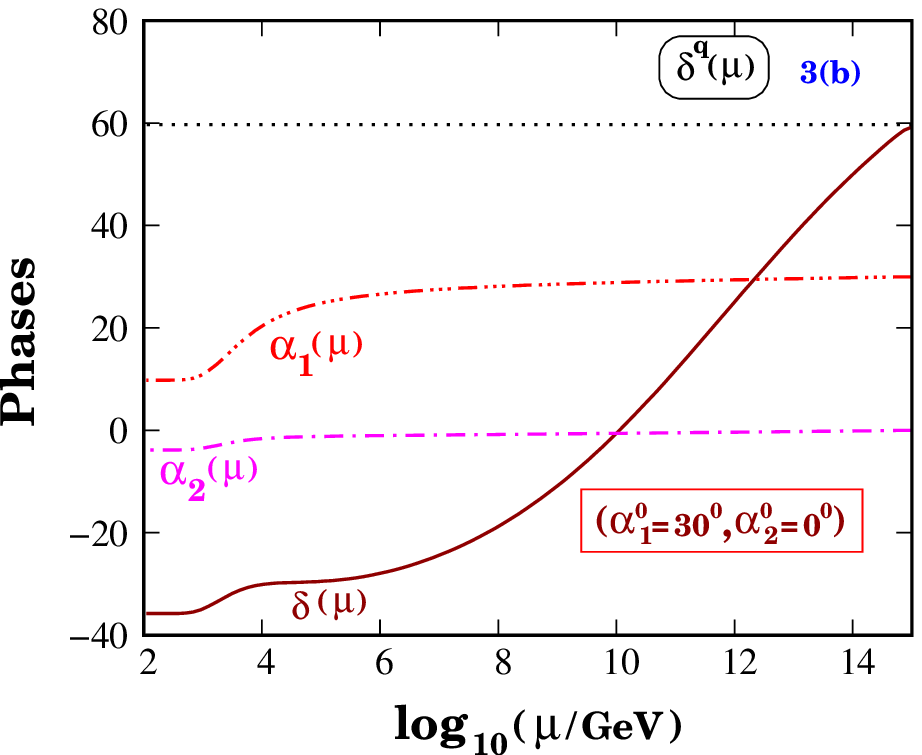}
  \caption{3(a).Prediction for the leptonic CP-violation parameter at low energies as a function of input Majorana phase $\ao^0$ at the GUT-seesaw scale.
3(b). Evolution of the leptonic Dirac phase and Majorana phases from the GUT-seesaw scale down to low energies. The horizontal dotted line shows the constancy of the CKM Dirac phase with unification point at the GUT-seesaw scale.}
  \label{jcp_phase}
\end{figure}

\begin{figure}[htbp]
  \centering  \includegraphics[width=0.45\textwidth,height=0.3\textheight]
  {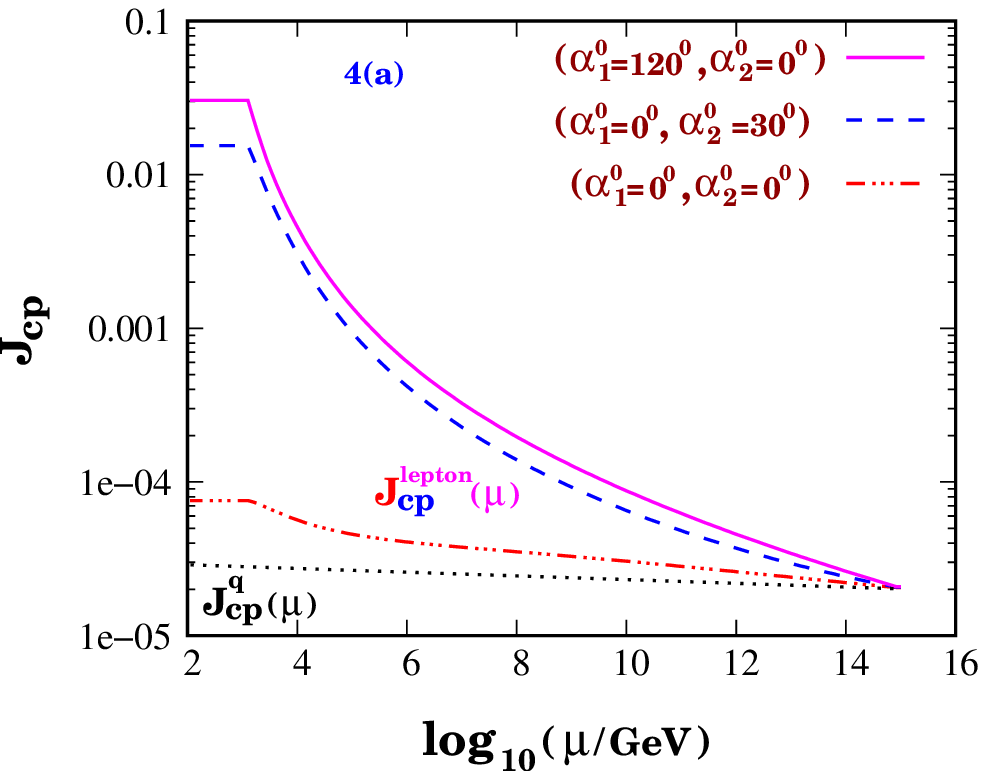}
  \includegraphics[width=0.45\textwidth,height=0.3\textheight]{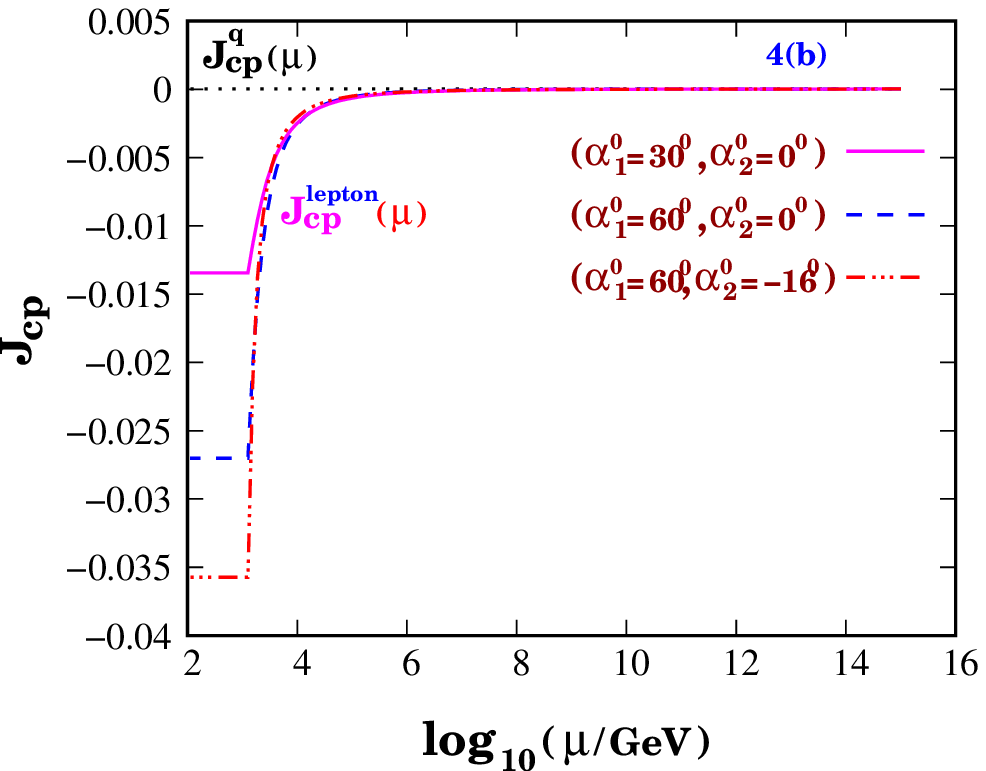}
  \caption{4(a). Evolution of the leptonic CP-violation parameter from the GUT-seesaw scale to low energies for different input values of  Majorana phases
and for positive values of $J_{CP}$.~Almost horizontal dotted line shows slow evolution of the corresponding
baryonic CP-violating parameter in the CKM matrix.
4(b). Same as Fig. 4(a) but only for negative values of the CP-violating parameter $J_{CP}$ obtained for different input values of Majorana phases.}
\label{jpos_neg}
\end{figure}

\begin{figure}[htbp]
  \centering  \includegraphics[width=0.45\textwidth,height=0.3\textheight]
  {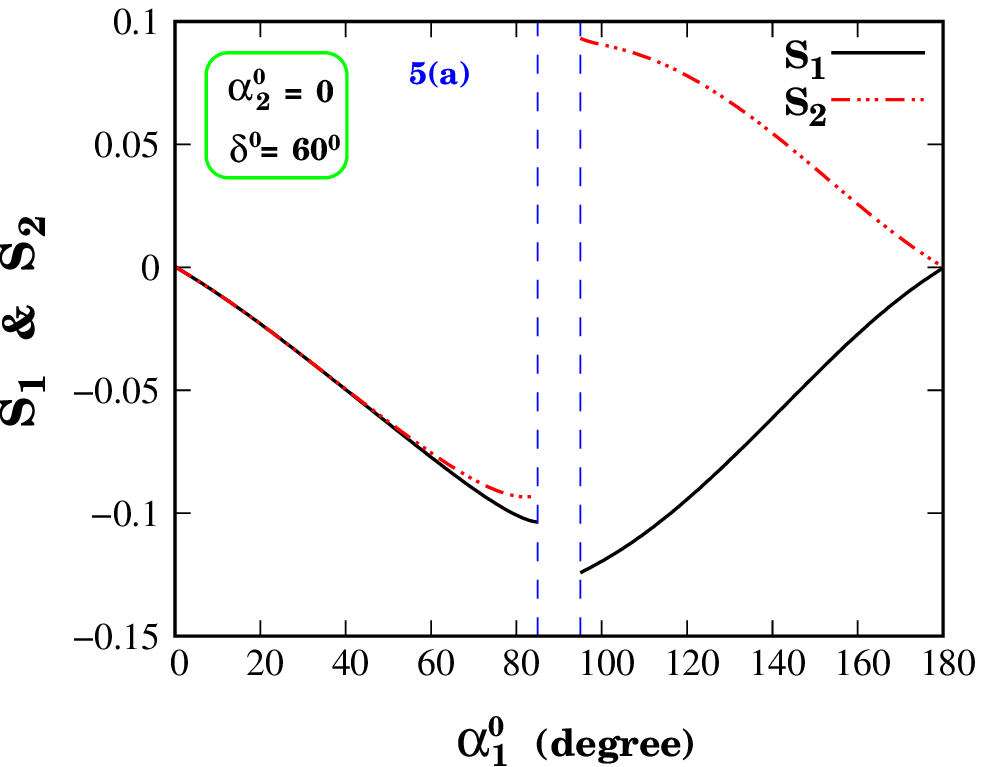}
\includegraphics[width=0.45\textwidth,height=0.3\textheight]{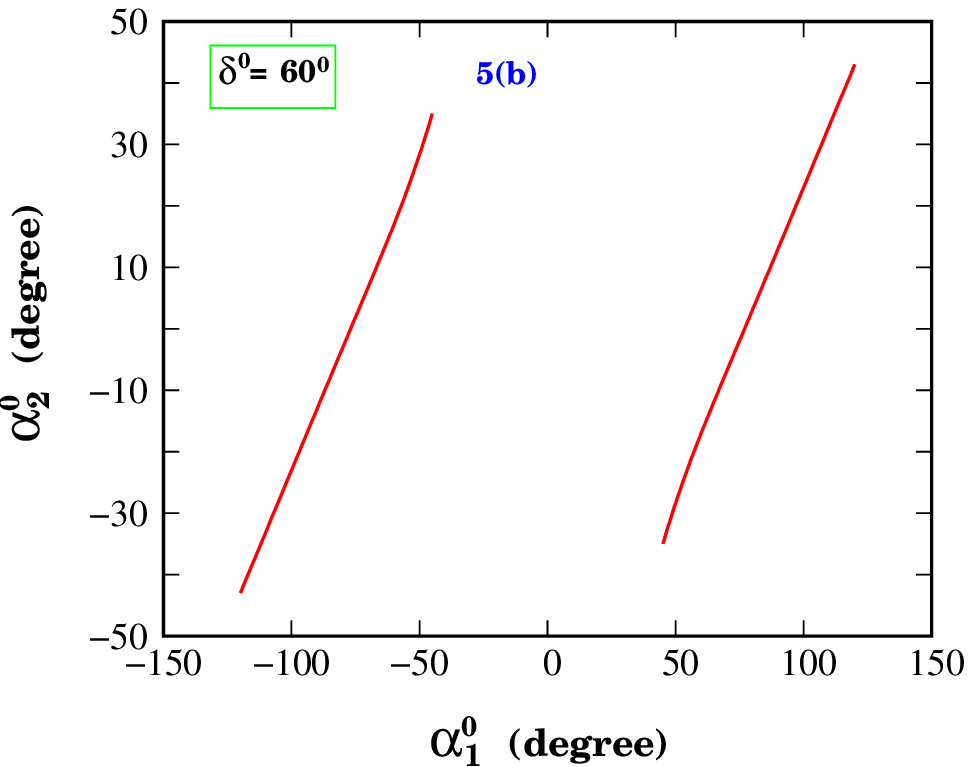}
\caption{5(a). Predictions of the two  invariants relating to  Majorana phases as defined in Sec.V.2.B in the text.
5(b). ~Two branches of the parameter space defined for no threshold corrections
through the magnification damping condition   
${\rm C}os ~2(\ao^0 - \at^0) = - 0.9$ at the 
GUT-seesaw scale.}
\label{s1s2_para}
\end{figure}

\begin{table*}
\caption{Solutions of RGEs for neutrino oscillation parameters with radiative magnification including Dirac and Majorana phases. The inputs for high scale mixings and the Dirac phase are from the CKM matrix as defined in the text. $M_{\tilde l}$ denotes the mass of slepton used for threshold correction for which we have used the wino mass $M_{\tilde w} = 150$ GeV. The GUT-seesaw scale is $M_R = 10^{15}$ GeV and $\tan\beta = 55$.}

\begin{ruledtabular}
\begin{tabular}{lcccccc}\hline
$\alpha^{0}_{1}$(deg)&40&60&140&45&120&0\\

$\alpha^{0}_{2}$(deg)&0&0&20&160&0&30\\

$m_1^0$(eV)&0.4969&0.4774&0.4975&0.4777&0.4774&0.497\\

$m_2^0$(eV)&0.5&0.48&0.5&0.48&0.48&0.5\\

$m_3^0$(eV)&0.577&0.554&0.577&0.554&0.554&0.577\\

$\alpha_{1}$(deg)&12.3195&17.6606&164.4949&15.4068&161.389&-8.22\\

$\alpha_{2}$(deg)&--5.1266&--6.686&8.4804&170.6779&6.29339&5.667\\

$\delta$(deg)&--43.513&--54.7767&76.7738&--70.0127&66.0524&41.2865\\

$m_1$(eV)&0.37055&0.356065&0.371006&0.356279&0.356044&0.3706\\

$m_2$(eV)&0.37128&0.356461&0.371422&0.35657&0.356476&0.37148\\

$m_3$(eV)&0.37467&0.359634&0.3745&0.359535&0.35964&0.37445\\

$(\Delta m^2_{21})_{\rm RG}$(eV$^2$)&$5.4339\times 10^{-4}$&$2.822\times 10^{-4}$&$3.0851\times 10^{-4}$&
$2.076\times 10^{-4}$&$3.0797\times 10^{-4}$&$6.346\times 10^{-4}$\\

$(\Delta m^2_{31})_{\rm RG}$(eV$^2$)&$3.048\times 10^{-3}$&$2.5544\times 10^{-3}$&$2.6092\times 10^{-3}$&
$2.3307\times 10^{-3}$&$2.5736\times 10^{-3}$&$2.849\times 10^{-3}$\\

$M_{\tilde e}/M_{\tilde \mu, \tilde \tau}$&1.51&1.39&1.42&1.49&1.39&2.82\\

$(\Delta m^2_{21})_{\rm th}$(eV$^2$)&$-4.6339\times 10^{-4}$&$-2.022\times 10^{-4}$&$-2.2851\times
10^{-4}$&$-1.1756\times 10^{-4}$&$-2.1796\times 10^{-4}$&$-5.546\times 10^{-4}$\\

$(\Delta m^2_{31})_{\rm th}$(eV$^2$)&$-0.9712\times 10^{-3}$&$-0.4179\times 10^{-3}$&$-0.4763\times 10^{-3}$&
$-0.2718\times 10^{-3}$&$-0.4185\times 10^{-3}$&$-0.96386\times 10^{-3}$\\

$\Delta m^2_{21}$(eV$^2$)&$8.0\times 10^{-5}$&$8.0\times 10^{-5}$&$8.0\times 10^{-5}$&
$9.0\times 10^{-5}$&$9.0\times 10^{-5}$&$8.0\times 10^{-5}$\\

$\Delta m^2_{31}$(eV$^2$)&$2.076\times 10^{-3}$&$2.1364\times 10^{-3}$&$2.1329\times 10^{-3}$&
$2.0589\times 10^{-3}$&$2.155\times 10^{-3}$&$1.88513\times 10^{-3}$\\

$\sin\theta_{12}$&0.5707&0.5497&0.5639&0.5828&0.5302&0.5417\\

$\sin\theta_{23}$&0.7211&0.7088&0.7145&0.7066&0.7092&0.722\\

$\sin\theta_{13}$&0.1177&0.1472&0.1512&0.1545&0.1518&0.1042\\

$J_{CP}$&-0.0187&-0.027&0.0335&-0.0335&0.0304&0.0155\\

$m_{ee}$(eV)&0.3524&0.322819&0.329399&0.314478&0.319534&0.359082\\

$m_{e}$(eV)&0.37084&0.356259&0.371215&0.356453&0.356246&0.370897\\\hline

\end{tabular}
\end{ruledtabular}
\label{tab1}
\end{table*}

\begin{table*}
\caption{Same as Table I. but without the necessity of threshold corrections.
The two input Majorana phases satisfy a strong damping condition
${\rm C}os ~2(\ao^0 - \at^0) = - 0.9$ ~which at the electroweak scale
 becomes  moderate with ${\rm C}os ~2(\ao - \at) \simeq  0.5 - 0.6$
for different solutions.}
\begin{ruledtabular}
\begin{tabular}{lcccccc}\hline
$\alpha^{0}_{1}$(deg)&45&60&65&75&90&120\\

$\alpha^{0}_{2}$(deg)&-35&-16&-12&-2&13&43\\

$m_1^0$(eV)&0.4781&0.4251&0.42504&0.43793&0.43795&0.49497\\

$m_2^0$(eV)&0.48&0.427&0.427&0.44&0.44&0.497\\

$m_3^0$(eV)&0.554&0.493&0.493&0.508&0.5079&0.573\\

$\alpha_{1}$(deg)&20.9358&16.558&17.687&20.4018&20.827&22.06\\

$\alpha_{2}$(deg)&-9.1444&-9.877&-9.259&-7.6403&-6.7838&-5.076\\
$\delta$(deg)&-81.3026&-74.553&-71.27&-63.403&-57.714&-49.246\\
${\rm C}os ~2(\ao - \at)$&0.497&0.603&0.589&0.558&0.570&0.580\\
$m_1$(eV)&0.356619&0.31708&0.32672&0.326665&0.32668&0.36921\\

$m_2$(eV)&0.356739&0.317208&0.32685&0.32679&0.32680&0.36933\\

$m_3$(eV)&0.3593271&0.319905&0.32967&0.329693&0.32961&0.37167\\

$\Delta m^2_{21}$(eV$^2$)&$8.56\times 10^{-5}$&$8.12\times 10^{-5}$&

$8.107\times 10^{-5}$&$8.17\times 10^{-5}$&$7.84\times 10^{-5}$&$8.86\times 10^{-5}$\\

$\Delta m^2_{31}$(eV$^2$)&$1.938\times 10^{-3}$&$1.8\times 10^{-3}$&$1.939\times 10^{-3}$&

$1.987\times 10^{-3}$&$1.92\times 10^{-3}$&$1.82\times 10^{-3}$\\

$\sin\theta_{12}$&0.4924&0.5865&0.572&0.5369&0.5501&0.5665\\

$\sin\theta_{23}$&0.700&0.69023&0.6912&0.69174&0.6982&0.7526\\

$\sin\theta_{13}$&0.165&0.1602&0.1609&0.16026&0.1608&0.1766\\

$J_{CP}$&-0.034&-0.03569&-0.0348&-0.03156&-0.0304&-0.02999\\

$m_{ee}$(eV)&0.306543&0.274384&0.283384&0.28645&0.288134&0.327606\\

$m_{e}$(eV)&0.356721&0.317195&0.326838&0.326778&0.326791&0.369324\\\hline

\end{tabular}
\end{ruledtabular}
\label{tab2}
\end{table*}

\section{ACKNOWLEDGMENTS}

\begin{acknowledgments}

M.K.P. thanks  Harish-Chandra Research Institute, Allahabad for hospitality
, Institute of Mathematical Sciences, Chennai for Senior Associateship, and
Institute of Physics, Bhubaneswar for research facility.
The work of R.N.M is supported by the NSF grant No.~PHY-0099544.
G.R. is supported by Raja Ramanna Fellowship of  DAE, Govt. of India.

\end{acknowledgments}


\begin{thebibliography}{99}


\bibitem{ps} J. C. Pati and A. Salam, Phys. Rev. {\bf D10}, 275
(1974).

\bibitem{georgi} H. Georgi and S. L. Glashow, Phys. Rev. Lett. {\bf 32}, 438
(1974).

\bibitem{so10} H. Georgi, Particles and Fields,{\it Proceedings of APS
Division of Particles and Fields,} ed C. Carlson, p575 (1975);
(1974); H. Frtzsch, P. Mikowski, Ann. Phys. {\bf 93}, 193 (1975).

\bibitem{rev} For a review and references, see
S. Raby, in Particle Data Group Book, W. -M. Yao {\it et
al.} J. Phys. {\bf G 33}, 1 (2006).

\bibitem{buras} M. Chanowitz, J. Ellis and M. K. Gaillard, Nucl.
Phys. {\bf B135}, 66 (1978); A. Buras, J. Ellis, M. K. Gaillard
and D. V. Nanopoulos, Nucl. Phys. {\bf B 135}, 66 (1978).

\bibitem{ananth} T. Banks, Nucl. Phys. {\bf B303}, 172 (1988);
M. Olechowski and S. Pokoroski, Phys. Lett. {\bf B214}, 393
(1988); B. Ananthnarayan, G. Lazaridis and Q. Shafi, Phys. Rev.
{\bf D 44}, 1613 (1991).

\bibitem{mpr1}R. N. Mohapatra, M. K. Parida and G. Rajasekaran, hep-ph/
0301234; Phys. Rev. {\bf D69}, 053007 (2004).

\bibitem{seesaw} P. Minkowski, Phys. Lett. {\bf B 67}, 421 (1977);
 M. Gell-Mann, P. Ramond and R. Slansky, in {\it
Supergravity},
eds. D. Freedman {\it et al.} (North-Holland, Amsterdam,
1980); T. Yanagida, in
proc. KEK workshop, 1979 (unpublished); R. N. Mohapatra and
G. Senjanovi\'c, Phys. Rev. Lett. {\bf 44}, 912 (1980); S. L. Glashow, in {\it
Proceedings of 1979 Cargese
Summer Institute on Quarks and Leptons}, eds. M. Levy {\it et al} , Plenum
 Press, New York, 1980, pp. 687-713.


\bibitem{mpr2} R. N. Mohapatra, M. K. Parida and G. Rajasekaran, Phys. Rev. {\bf D 71}, 057301 (2005).


\bibitem{maltoni} M. Maltoni, T. Schwetz, M. Tortola and J. F. Valle, arXiv:
hep-ph/0405172; New J. Phys. {\bf 6}, 122 (2004).

\bibitem{chooz}  M.~Appollonio {\it et~al.},~Phys.~Lett.~{\bf B466}, 415 (1999);
~F. Boehm {\it~et~al.}, Phys. Rev. {\bf D64}, 112001 (2001).


\bibitem{bb} For an overview of the present experimental program, see A.~Barabash,
  arXiv: hep-ex/0608054; Invited talk at the {\it Neutrino 2006}
  conference, Santa Fe (2006).



\bibitem{klapdor} H. V. Klapdor-Kleingrothaus {\it et al.} Eur. Phys. J. {\bf A 12},
147 (2001); C. Aalseth {\it et al.} Phys. Rev. {\bf D 65}, 092007 (2002);
H. V. Klapdor-Kleingrothaus {\it et al.}, Mod. Phys. Lett. {\bf
A 16}, 2409 (2001); hep-ph/0303217;
 H. V. Klapdor-Kleingrothaus, A. Dietz and I. V. Krivosheina,
Foundations of Physics {\bf 32}, 1181 (2002).

\bibitem{cremo} For a review, see O. Cremonesi, Nucl. Phys.  (Proc. Suppl.)
{\bf B  118}, 287 (2003).


\bibitem{katrin} A. Osipowicz {\it et~al.}, (KATRIN Project), hep-ex/0109033.

\bibitem{wmap} D. Spergel {\it et al.} astro-ph/0302209; astro-ph/0603449:  C. L. Bennett {\it et al.}
astro-ph/0306207; S. Hannestad, JCAP,{\bf 0305}, 004 (2003);
astro-ph/0303076; J. R. Kristiansen, O. Elgaroy and  H. K.
Eriksen, astro-ph/0608017.

\bibitem{hannestad} S. Hannestad, Phys. Rev. Lett. {\bf 95}, 221301 (2005);

\bibitem{fukugita}  M. Fukugita, K. Ichikawa, M. Kawasaki and O. Lahav, Phys. Rev. {\bf D 74}, 027302 (2006);
M. Fukugita, Nucl. Phys. (Proc. Suppl.) {\bf B 155}, 10 (20006);
M. Tegmark {\it et al.} Phys. Rev. {\bf D 69}, 103501 (20004).


\bibitem{smir} B.~Kayser, Phys.\ Rev.\ D {\bf 30}, 1023 (1984);
For a recent discussion, see Y. Farzan and A. Yu Smirnov,
hep-ph/0610337.

\bibitem{wett} C.Wetterich,
Nucl. Phys. {\bf B 187}, 343 (1981); A. Ioannisian, Sov. J. Nucl. Phys. 
{\bf 51}, 511 (1990).
\bibitem{babu}K. S.~Babu, C. N.~Leung and J.~Pantaleone, Phys.~Lett.~{\bf
B319}, 191 (1993); 
P. Chankowski and Z. Pluciennik, Phys. Lett. {\bf B316},312 
 (1993);   
M. Tanimoto, Phys. Lett. {\bf B 360}, 41 (1995); 
J. Ellis and S. Lola, Phys. Lett. {\bf B 458}, 310 (1999); E. Ma,
J. Phys. {\bf G25}, L97 (1999); N. Haba, Y. Matsui, and N. Okamura,
Eur. Phys. J. {\bf C 17}, 513 (2000);   
S. Antusch, M. Drees, J. Kersten, M. Lindner and M. Ratz,
Phys. Lett. {\bf 519}, 238 (2001); Phys. Lett. {\bf 525}, 130 (2002);
 G. Bhattacharyya, A. Raychaudhuri and A. Sil, Phys.
Rev. {\bf D 67}, 073004 (2003); A. S. Joshipura, S. D. Rindani and N. N. Singh
Nucl. Phys. {\bf B 660}, 362 (2003); A. Dighe,
 S. Goswami and  P. Roy, Phys. Rev. {\bf D 73}, 071301 (2006)


\bibitem{bdmp} K. R. S.~Balaji, A. S.~Dighe, R. N.~Mohapatra and M. K.~Parida,
Phys.~Rev.~Lett.~{\bf 84}, 5034 (2000); Phys.~Lett.~{\bf B481}, 33 ~(2000);
K. R. S.~Balaji,  R. N.~Mohapatra, M. K.~Parida and ~E. A. Paschos,
Phys.~Rev.{\bf D 63}, 113002~(2001).

\bibitem{casas}P. H.~Chankowski, W.~Kr\'{o}likowski and S.~Pokorski,
Phys. Lett. {\bf B 473}, 109 (2000); J. A.~Casas, J. R.~Espinosa, A.~Ibarra 
and I. Navarro, ~Nucl. Phys. {\bf B 569}, 82 (2000); ~hep-ph/9910420; 
P. H. Chankowski and S. Pokorski,
Int. J. Mod. Phys. {\bf A 17}, 575 (2002). 

\bibitem{antusch1} S. Antusch, J. Kersten, M. Lindner and M. Ratz,  Phys. Lett. {\bf B 544}, 1 (2002);
 S. Antusch, P. Huber, J. Kersten, T. Schwetz,
 and W. Winter, arXiv: hep-ph/0404268; T. Miura, T. Shindou and E. Takasugi,
 arXiv: hep-ph/0308109 ; T. Shindou and  E. Takasugi, arXiv: hep-ph/0402106.

\bibitem{antusch2}  S. Antusch,  J. Kersten, M. Lindner and M. Ratz, Nucl. 
Phys. {\bf B 674}, 401 (2003);  S. Antusch,  J. Kersten, M. Lindner, M. Ratz
and  M. A. Schmidt, JHEP {\bf 0503}, 024 (2005).

\bibitem{ellis} J. Ellis, A. Hektor, M. Kadastik, K. Kannike and  M. Raidal,
Phys. Lett. {\bf B 631}, (2005) 32.


\bibitem{xing} S. Luo, J. Mei and  Z. Xing, Phys. Rev. {\bf D 72},
053014 (2005);
J. Mei, arXiv: hep-ph/0502015 ; Z. Xing, arXiv: hep-ph/0510312; S. Luo
and  Z. Xing, arXiv: hep-ph/0603091.

\bibitem{mpr3} R. N. Mohapatra, M. K. Parida and G. Rajasekaran, Phys. Rev. {\bf D 72}, 013002 (2005).

\bibitem{lee} D. G. Lee and R. N. Mohapatra, Phys. Lett. {\bf B 329},
463 (1994); C. Hagedorn, M. Lindner and R. N. Mohapatra, JHEP,
{\bf 0606}, 042 (2006).

\bibitem{falk} N. K. Falk. Z. Phys. {\bf C 30}, 247 (1986); K. S. Babu, Z. 
Phys. {\bf C 35}, 69 (1987); S. G. Neculich, Phys. Rev. {\bf D 48}, 5293
(1993); P. Binetray and P. Ramond, Phys. Lett. {\bf B 350}, 49 (1995);
K. Wang, Phys. Rev. {\bf D 54}, 3513 (1994). 


\bibitem{pdg} W. -M. Yao {\it et al.}, J. Phys. {\bf G 33}, 1 (2006).



\bibitem{jarlskog} C. Jarlskog, Phys. Rev. Lett. {\bf 55}, 1039 (1985).

\bibitem{up} V. Barger, M. S. Berger, and P. Ohmann, Phys. Rev. {\bf D47},
1093 (1993); C. R. Das and M. K. Parida, Eur. Phys. J. {\bf C 20}, 121
(2001).

\bibitem{chun} 
 P. Chankowski and P. Wasowicz, Eur. Phys. J. {\bf C23}, 249
(2002);  E. J. Chun and S. Pokorski, Phys. Rev. {\bf D62}, 053001 (2000).

\bibitem{rnm} R. N. Mohapatra {\it et al.} arXiv: hep-ph/0412099, R. N. Mohapatra {\it et al.} arXiv: hep-ph/0510213.

\bibitem{picariello} J. A. Aguilar-Saavedra and G. C. Branco, Phys. Rev. 
{\bf D 62}, 096009 (2000); M. Picariello, arXiv: hep-ph/0611189.

\end{thebibliography}
\end{document}